\documentclass[pdflatex,sn-mathphys-num]{sn-jnl}

\usepackage{appendix}
\usepackage{graphicx}%
\usepackage{multirow}%
\usepackage{amsmath,amssymb,amsfonts}%
\usepackage{amsthm}%
\usepackage{mathrsfs}%
\usepackage[title]{appendix}%
\usepackage{xcolor}%
\usepackage{textcomp}%
\usepackage{manyfoot}%
\usepackage{booktabs}%
\usepackage{algorithm}%
\usepackage{algorithmicx}%
\usepackage{algpseudocode}%
\usepackage{listings}%
\usepackage{subcaption}
\theoremstyle{thmstyleone}%
\theoremstyle{thmstyletwo}%

\theoremstyle{thmstylethree}%

\begin{document}

\title[Article Title]{Revisiting the Reported Period of FRB 20201124A Using MCMC Methods}


\author[1]{\fnm{Jun-Yi} \sur{Shen}}

\author[1]{\fnm{Yuan-Chuan}\sur{Zou}}\email{zouyc@hust.edu.cn}

\affil*[1]{\orgdiv{Department of astronomy, School of Physics}, \orgname{Huazhong University of Science and Technology}, \orgaddress{\street{Luoyu road}, \city{Wuhan}, \postcode{430074}, \state{Hubei}, \country{China}}}


\abstract{Fast radio bursts (FRBs) are millisecond-duration radio transients whose physical origin remains uncertain. Magnetar-based models, motivated by observed properties such as polarization and large rotation measures, suggest that FRB emission may be modulated by the magnetar spin period. We present an efficient method to search for periodic signals in repeating FRBs by combining phase folding and Markov Chain Monte Carlo (MCMC) parameter estimation. Our method accelerates period searches. We test the method using observational data from repeater FRB 20201124A, and show that it can recover reported candidate periods.}




\maketitle
\section{Introduction}
Fast radio bursts (FRBs) are millisecond-duration, high
luminosity (typically $\ge 10^{39}$ erg $\rm{s^{-1}}$) extremely high brightness temperature radio transients, first reported by Lorimer in 2007 \cite{2007Sci...318..777L}.  
These radio signals are scattered by plasmas and  exhibit dispersion measures (DMs), strongly suggesting that FRBs originate from cosmological distances. 
The limited spatial resolution of previous radio telescopes made localizing non-repeating FRBs challenging, a problem largely alleviated by the advent of CHIME and KKO, which provide much higher spatial resolution \cite{2025ApJS..280....6C}.
However, some FRBs are repeaters, allowing their host galaxies to be securely identified through multi-wavelength observations. 
In addition, repeating FRBs are monitored over much longer observational timescales than non-repeating FRBs, resulting in substantially more data and studies of repeaters.
The first known repeating FRB is FRB 20121102 \cite{2016Natur.531..202S}, which was subsequently localized by \citet{2017Natur.541...58C}.
Large samples of repeating FRBs make it possible to investigate their activity patterns, repetition rates, and energy distributions.

The physical origin of FRBs remains unclear. Magnetar models have been proposed to explain their observational properties, such as polarization and large rotation measures \cite{2014MNRAS.442L...9L, 2017ApJ...843L..26B, 2020ApJ...899L..27M}. 
In particular, a milestone observation supporting these models was the discovery of FRB 20200428 \cite{2020Natur.587...54C}, which was associated with an X-ray burst from the Galactic magnetar SGR J1935+2154 \cite{2020Natur.587...59B}. 
The association between FRBs and magnetars suggests that radio emission from FRBs may be modulated by the spin period of the magnetar, typically on second-long timescales. This possibility has attracted considerable attention in recent years. 
Several studies have investigated this possibility (e.g., \cite{2022RAA....22l4004N, 2024ApJ...977..129D, 2025arXiv250312013D}), but so far no compelling evidence has been found.
Apart from \citet{2025arXiv250312013D}, which reported a candidate periodicity of  1.706015 s on MJD
59310 and 1.707972 s on MJD 59347 for FRB 20201124A, the majority of active windows for this source do not exhibit this $\sim$ 1.7 s periodic behavior.
Furthermore, \cite{2025arXiv250514219G} raised a contrary view, suggesting that this may not be a genuine period. 
Motivated by this, we also conducted a study on the periodicity of FRBs.
Standard methods for searching for periodicities in astronomical sources include Epoch Folding (EF), the H-test, Fast Fourier Transform (FFT), the Lomb-Scargle periodogram (LSP), and others; see, e.g., \citet{1983ApJ...266..160L, 2021ApJ...909...33B, 2019ApJ...881...39H, 2018ApJS..236...16V}.

These methods can also be applied to search for periodicities in FRBs. 
In this work, we apply the Monte Carlo Markov Chain (MCMC) method together with the EF and show that MCMC provides a rough estimate of the period.

This article is organized as follows. In Section 2, we describe our method, which can accelerate the period search. Section 3 demonstrates the applicability of the method, and in Section 4, we discuss our findings.

\section{Method} \label{sec:style}

Several methods are available for studying periodicity in astronomical data. Different astronomical objects exhibit different observational characteristics and therefore require different methods for periodicity analysis. 
The FFT decomposes the time series into sinusoidal components of different frequencies, and the dominant period is determined from the frequency with the highest spectral power.
The Lomb–Scargle periodogram (LSP) is analogous to the Fourier transform but is specifically designed to analyze unevenly sampled data. It is commonly used to search for sinusoidal periodicities by fitting sinusoids using a least-squares approach. 
Therefore, the LSP is more suitable for signals with good phase coverage within each period.
For instance, the X-ray emission from SGRs is consistent with radiation from a large emitting area and shows a blackbody component in spectral analyzes (see, e.g., \citet{2003MNRAS.343.1301P}). 
The emission originates from an extended region and is radiated over a large solid angle. As a result, over the SGR spin period, the observed emission exhibits relatively smooth variability and a high duty cycle ($\geq 10\%$). 
Unlike SGRs, the light curve of the pulsar shows narrow pulsations and a low duty cycle (1$\%-5\%$; see Chapter 10 of \citet{1983bhwd.book.....S}).
The expected brightness temperature of an FRB is so high that the pulse cannot be emitted from a black body.
FRB radiation is distinct from isotropic blackbody emission and may be strongly beamed. 
We assume that the duty cycle of FRBs is comparable to that of pulsars. Consequently, methods used in pulsar searches can be applied to detect periodicity in FRBs. 
Especially, even if the actual duty cycle of the FRBs differs from our assumption, the method remains applicable for searching periodicities. For sources with low duty cycles, such as pulsars or highly pulsed FRBs, EF provides a particularly effective approach.

The EF and H-tests focus on detecting phase clustering in the data. Following \citet{1983ApJ...266..160L}, data covering a duration $T$ are folded using a trial period $P$ into $n$ phase bins, resulting in the folded pulse profile. We apply the EF method to search for periodicities in FRBs by identifying the trial period $P$ that maximizes phase concentration, resulting in a corresponding peak in the folded phase profile.
The full phase range $[0, 2\pi)$ is divided into $N$ bins of width $\Delta \phi = 2\pi/N$. The $j$-th bin covers the interval $[\phi_j, \phi_{j+1})$ with $\phi_j = j \Delta \phi$ for $j=0,1,\dots,N-1$. For a trial period $P$ close to the true period, the initial phases of the TOAs, $\phi_i = \mathrm{MOD}(t_i / P)$, should be clustered.
$O_j$ denotes the number of pulses that fall within the range $[\phi_j, \phi_{j+1})$. Under the null hypothesis, the counts in each bin follow a Poisson distribution.
For sufficiently large expected counts, the $\chi^2$ statistic,
\begin{equation}
    \chi^2_{N} = \sum_{j=1}^{N} \frac{(O_j - E_j)^2}{\sigma_j^2},
\end{equation}
approximately follows a $\chi^2$ distribution with $N$ degrees of freedom,
where $O_j$ and $E_j$ are the observed and expected counts in the $j$th bin.
The $\sigma_j$ is a weighting factor related to the ratio of the total observation integration time of the j-th pulse phase bin versus the total observation time; see \citet{1983ApJ...266..160L}. 
A reduced $\chi^2/N \simeq 1$ suggests the absence of statistically significant
pulsations in the data.
The significance of a candidate signal is assessed through the corresponding
$p$-value obtained from the $\chi^2$ distribution, which represents the
probability that the observed deviations are due to random fluctuations.

\begin{figure}[h] 
    \centering
    \includegraphics[width=0.4\textwidth]{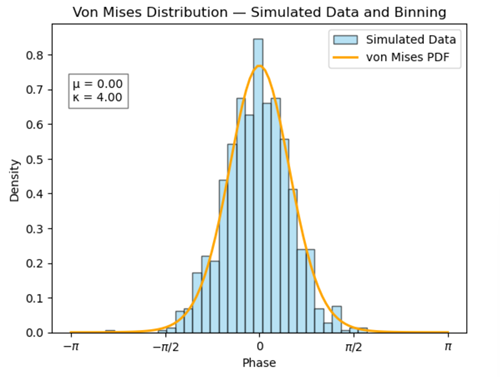} 
    \caption{Schematic diagram of a Von Mises distribution. The simulated data represent a concentrated phase profile, which can be well described by a Von Mises distribution. Therefore, we adopt the Von Mises distribution to model the optimal profile.}
    \label{fig:VMD}
\end{figure}

The objective is to identify a trial period $P$ that produces a highly
concentrated folded phase distribution, resulting in a peak in the phase profile.
The peak is modeled using a parametric Von Mises distribution, which serves as the circular analog of a Gaussian distribution. Its probability density function is given by:
\begin{equation}
f(\theta \mid \mu, \kappa)
= \frac{1}{2\pi I_0(\kappa)} 
  \exp\!\left[ \kappa \cos(\theta - \mu) \right].
\end{equation}
The parameters $\mu ,\kappa $ represent the mean phase and the concentration of the distribution, respectively.
As shown in Fig. \ref{fig:VMD}, the Von Mises distribution can be applied to assess the quality of a trial period $P$.
Given a trial period $P$, times of arrival (TOAs) are converted into rotational phases,
$\{\phi_1, \phi_2, \ldots, \phi_n\}$.
We assume that the phase distribution follows a Von Mises distribution.
The corresponding log-likelihood function is therefore given by

\begin{equation}
\ln \mathcal{L}
= \sum_{j=1}^{N} O_j
\ln \left[
f(\phi_j^* \mid \mu, \kappa)\
\right] ,
\end{equation}
Here, $\phi_j^{*}$ denotes the central phase of the $j$-th phase bin, defined as
$\phi_j^{*} = (\phi_j + \phi_{j+1})/2$.
Maximizing this likelihood yields the best-fitting parameters, corresponding to the most strongly concentrated phase distribution and thus the optimal trial period.

\section{Data}
Du et al. reported a candidate periodicity on a timescale of seconds in FRB 20201124A \cite{2025arXiv250312013D}. They reported a period of $1.706015(2) s$ at MJD 59310 and $1.707972(1) s$ at MJD 59347. 
FRB 20201124A was first discovered by CHIME \cite{2021ATel14497....1C}. FRB 20201124A is the third FRB source associated with a persistent radio source (PRS) \cite{2024Natur.632.1014B}. The presence of an associated PRS provides constraints on the nature of FRBs \cite{2023ApJ...958..185C}. 
We downloaded data from the FAST public archive, \footnote{\url{https://psr.pku.edu.cn/index.php/publications/frb20201124a/}} , covering observations from MJD 59307 to MJD 59360 with a total observing time of 91 h. \cite{2022Natur.609..685X}. The daily event rate varies between $6 \pm 1 h^{-1}$ and $ 46 \pm 8h^{-1}$. The data were recorded with a frequency resolution of 122.07 kHz and 
a temporal resolution of 49.152 $\mu$s or 196.608 $\mu$s. The burst search was performed using the TransientX software, with a signal-to-noise (S/N) ratio threshold of $\geq 7$.\footnote{\url{https://github.com/ypmen/TransientX}}  In addition, \citet{2022RAA....22l4004N} searched for periodicity in this source using FAST observations from MJD 59482 to MJD 59485, corresponding to a highly active window.

\section{Result}
As discussed in Section 2, we perform a Markov Chain Monte Carlo (MCMC) analysis. We use the \texttt{emcee} package \citep{2013PASP..125..306F}, which is widely applied in astronomy.
Before parameter estimation, the long dataset is divided into several segments for two reasons. 
First, the search resolution should be chosen so that each step satisfies $\delta P \cdot T/P^2 \le 0.1 $, which does not cause the profile to change significantly. Shorter datasets are faster to compute, effectively resulting in a coarser search resolution.
Second, the influence of $\dot P$ can be ignored. For long-duration observations, the phase is given by
$\phi_{0,i} =\mathrm{MOD}[{t_i}/{P} - 0.5 \, \dot{P} \, {t_i^{2}}/{P^{2}} ]$.
However, short data segments may lead to spurious periodicities. The false-alarm probability (FAP) can be quantified using statistical methods. For example, the $\chi^{2}$ statistic can be used to estimate the probability that the observed phase concentration arises by chance.
Another criterion for a genuine period is its consistent detection in most MJD data segments. 
We perform the following steps:
\begin{enumerate}
\item The FAST observations of FRB 20201124A, spanning MJD 59307 to MJD 59360, were divided into daily segments. After excluding days without detected bursts from FRB 20201124A, the remaining data consist of 44 days with burst detections.
\item Parameter estimation using MCMC is performed to identify the folded profile ${O_j}$ that best fits a von Mises distribution, which is adopted to model the phase profile.  All candidates merit further study.
\item For each data segment, the MCMC analysis provides a $1\sigma$ confidence interval for the period. We then search for periods within this interval at sufficiently fine resolution and identify the period corresponding to the maximum $\chi^2$, which is taken as the final candidate for the current data segment.
\item The final step is to verify whether the estimated period $P$ is consistent across different daily segments. The true period should be present in most MJD segments. This allows the confidence level of each candidate to be assessed, and a reliable conclusion to be drawn. 
\end{enumerate}
\subsection{review of $\sim 1.7$ s period of FRB 20201124A}
\begin{figure}[h]
    \centering

    \begin{subfigure}{0.4\textwidth}
        \centering
        \includegraphics[width=\textwidth]{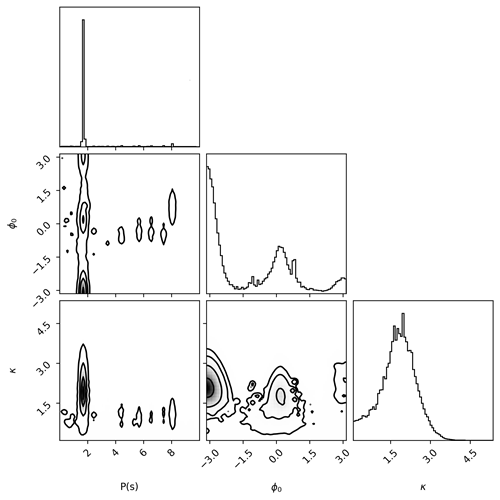}
        \caption{MCMC corner plot of data segment of MJD 59310}
        \label{fig:59310_sub}
    \end{subfigure}
    \begin{subfigure}{0.4\textwidth}
        \centering
        \includegraphics[width=\textwidth]{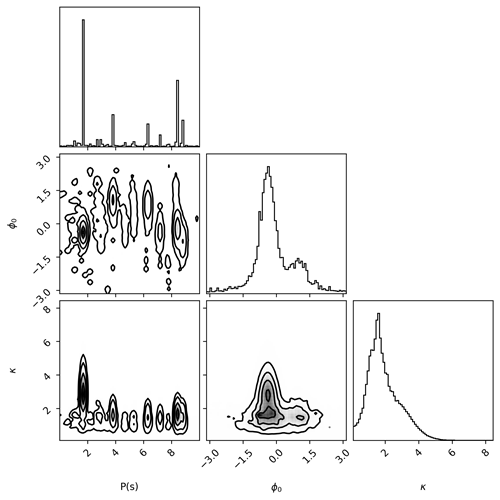}
        \caption{MCMC corner plot of data segment of MJD 59347}
        \label{fig:59310_sub}
    \end{subfigure}

    \caption{ (a) Parameter estimation was performed based on FAST observations on MJD 59310. The primary peak of the period $P$ occurs at 1.680344–1.706105 s ($1\sigma$). The parameter $\phi_0$ exhibits a secondary peak around 0, which may correspond to a minor peak in $P$. (b) Data segment from MJD 59347. Several minor peaks of $P$ are present, which we attribute to the short duration of the data causes a multiple peaks. The major peak lies at 1.707917–1.707957 s ($1\sigma$). These two period search results are consistent with \citet{2025arXiv250312013D}.}
    \label{fig:59310and59347}
\end{figure}

In this section, we test the applicability of our method. Our approach performs MCMC parameter estimation, which is implemented without a predefined search step. Determining the period with very high precision is beyond the capability of our method.
The resolution of the period search depends on the observational duration $T$ of the data segment.
The resolution of $P$ should satisfy $\delta P \cdot T/P^2 \le 0.1$ . If the required resolution is finer than the search step, the MCMC cannot locate the maximum likelihood, resulting in a chain that fails to converge. We applied our method as described above. 
We set the parameters $[P, \phi_0, \kappa]$ as initial [1,0,1], twenty-five walkers, and $10^5$ total steps. The prior distributions are $P \in  (0,10)$ s, $\phi_0 \in (- \pi, \pi)$, and $\kappa \in (0,10)$.
The resulting MCMC corner plots are shown in Fig. \ref{fig:59310and59347}. These figures show the best estimates of the parameter $P$. 
As shown in Fig. \ref{fig:59310and59347}, our method is capable of recovering the previously reported $\sim$ 1.7 s candidate. The next step is to apply this method to search for additional possible periods.
\subsection{other result}
\begin{figure}[ht]
    \centering
    \includegraphics[width=0.3\textwidth]{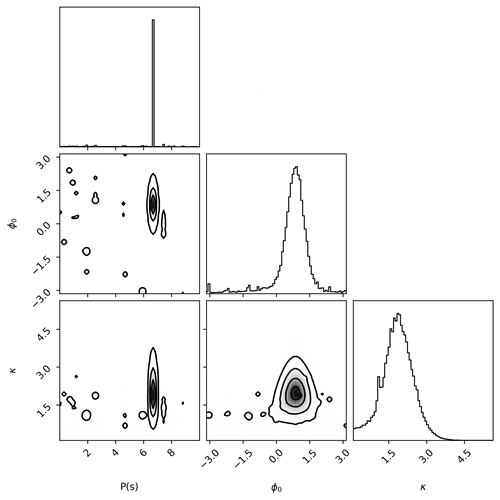}\textbf{(a)}
    \includegraphics[width=0.3\textwidth]{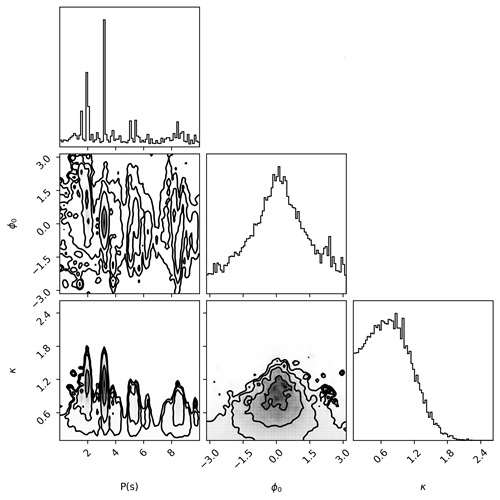}\textbf{(b)}
    \includegraphics[width=0.3\textwidth]{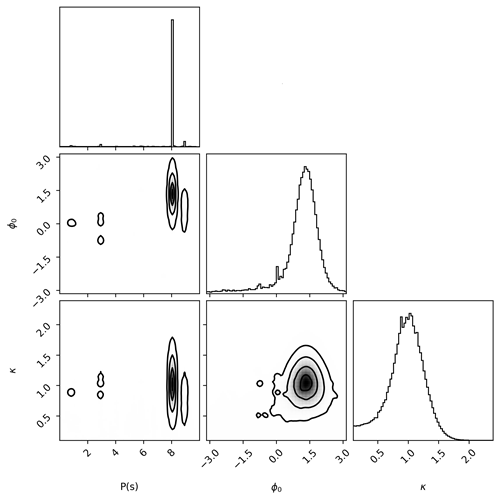}\textbf{(c)}
    \includegraphics[width=0.3\textwidth]{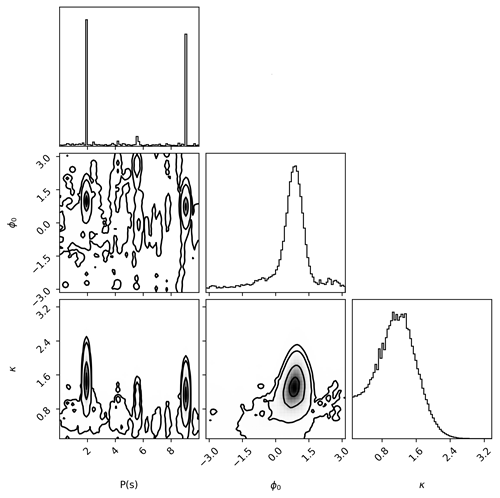}\textbf{(d)}
    \includegraphics[width=0.3\textwidth]{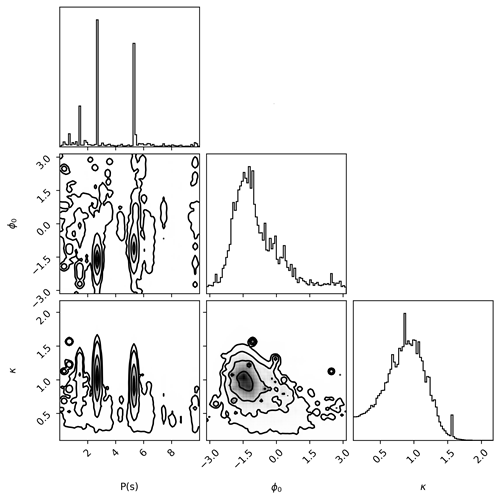}\textbf{(e)}
    \includegraphics[width=0.3\textwidth]{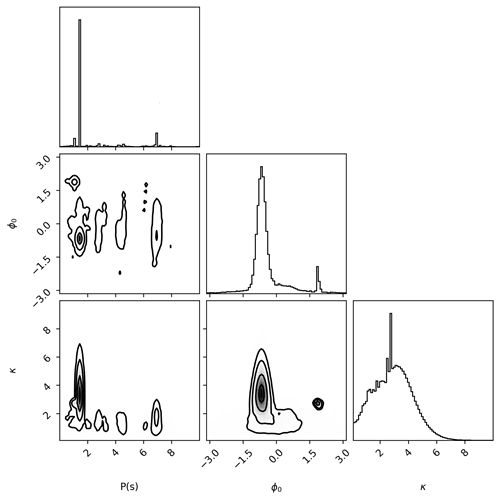}\textbf{(f)}
    \caption{
    \textbf{data segment:}
    (a) MJD 59309,
    (b) MJD 59311,
    (c) MJD 59318,
    (d) MJD 59329.
    (e) MJD 59337,
    (f) MJD 59339.
    This figure shows representative MCMC results from our study. Certain corner plots, shown in (a), (c), and (f), exhibit good convergence. In contrast, the corner plot in (b) appears highly chaotic, possibly due to either a lack of true periodicity in the data or the short data segment, which can easily produce spurious period detections.
    }
    \label{fig:combined}
\end{figure}

Here, we present several examples of the MCMC results from different data segments, as shown in Fig. \ref{fig:combined}.
We present a table of candidate major and minor peaks identified in Table \ref{tab:candidate_periods}. Most of these candidates are above 3$\sigma$ by $\chi^2$ statistics. However, we find no consistency among these candidates.

\begin{figure}[h] 
    \centering
    \includegraphics[width=0.5\textwidth]{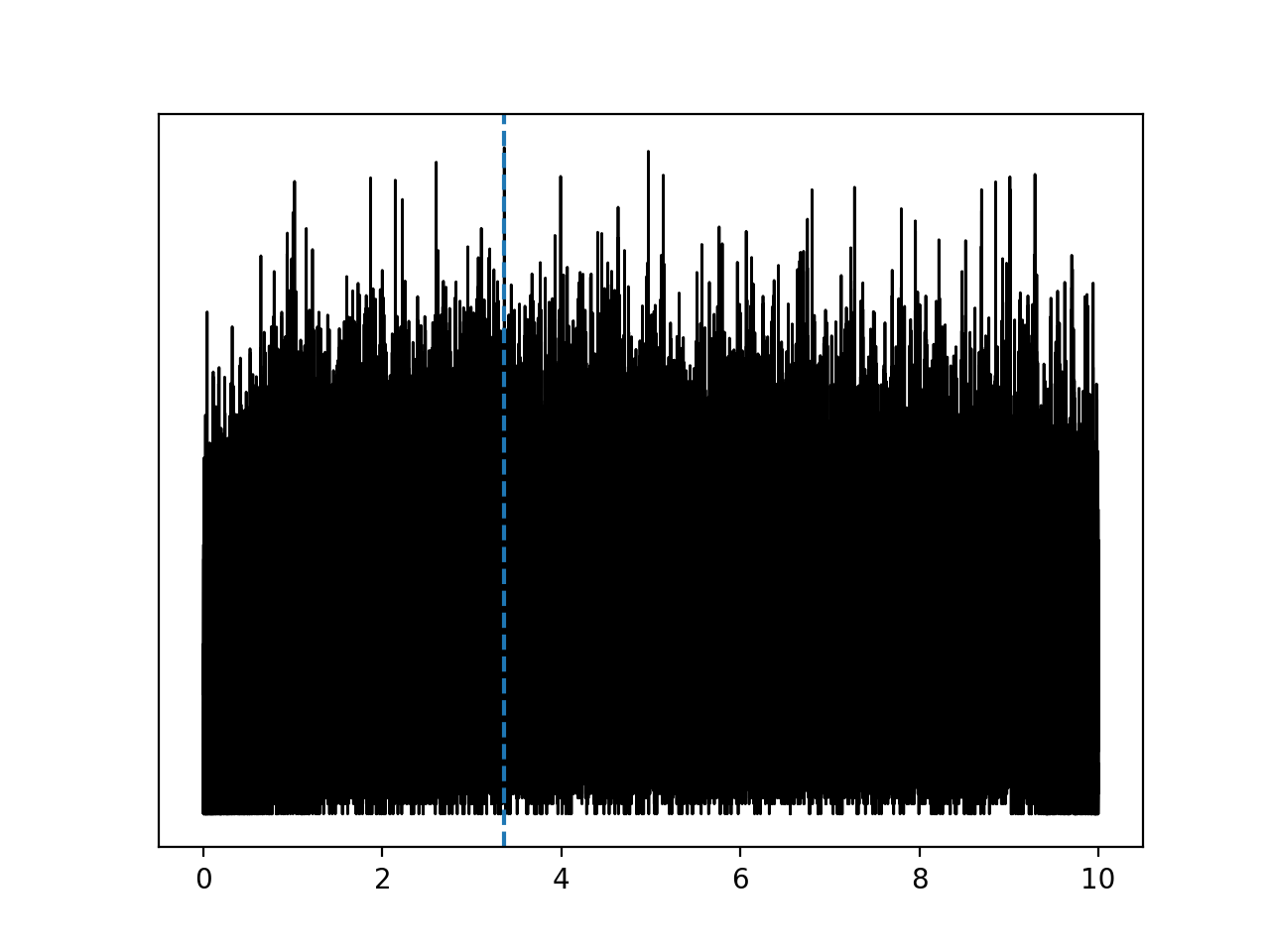} 
    \caption{The normalized plot of the period search over (0,10) s shows no evidence of a preferred period. The dotted line indicates the location of the maximum count, which does not reveal any significant or special value.}
    \label{fig:hist_para}
\end{figure}

A periodic signal with low individual significance can be revealed by summing the confidence levels over all segments. We therefore search for a preferred value of the parameter $P$ over the data from  MJD 59307 to MJD
59360. The prior range of $P \in (0,10)$ s, is divided into 500,000 bins. The number of counts in each bin is used as a proxy for significance. To suppress the influence of extreme values from individual data segments, we adopt the logarithm of the counts.
The logarithms of the sample counts across all segments from MJD 59307 to MJD 59360 are then summed to detect weak periodic signals present across multiple data segments, as shown in Fig. \ref{fig:hist_para}.

\section{Conclusion and Discussion}
In this work, we presented a novel and efficient method for detecting periodic signals in repeating Fast Radio Bursts (FRBs) by combining Phase Folding with Markov Chain Monte Carlo (MCMC) parameter estimation. By modeling the folded phase profile with a Von Mises distribution, our approach effectively targets sources with low duty cycles, a characteristic shared by pulsars and FRBs. We applied this pipeline to FAST observations of FRB 20201124A to test its validity and revisit previously reported periodicities.

We draw the following conclusion: Firstly, the $\sim$ 1.7s period can be found in MJD 59310 and MJD 59347 of FRB 20201124A. However, in the other dataset, we did not find a periodic signal near 1.7 s. Secondly, our new method can be used to perform a coarse search for the period, serving as a computationally efficient alternative to iterating over all parameters \cite{2025arXiv250312013D}, while still providing a good search direction through the use of MCMC. In addition, the data can be processed in parallel, which accelerates the computation.

Fast Radio Bursts (FRBs) are enigmatic astronomical phenomena. A milestone in FRB observation is FRB 20200428, which was associated with a magnetar in the Milky Way \cite{2020Natur.587...59B}. 
SGRs are magnetars, a type of neutron star that exhibits spin-period-like behavior during active bursts (see, e.g., \citet{1999ApJ...510L.115K}). 
Ultra luminous X-ray sources (ULXs) were discovered with orbital and spin periods by \citet{2014Natur.514..202B}. Until recently, the main problem with unveiling the nature of the ULXs stemmed from the absence of reliable measurements of the masses of the accreting compact objects. 
This has changed with the discovery of pulsing ultra-luminous X-ray sources.
The detection of a period could provide important constraints on the central engine of FRBs. However, many studies have so far failed to find such periodicities.
Discovering a periodicity in FRB would represent a significant milestone in FRB research; however, despite numerous efforts, no conclusive results have yet been reported.

We believe that FRBs are similar to pulsars in exhibiting low duty cycles. Therefore, we adopted the EF method, which is commonly applied in pulsar searches. We re-examine the 1.7 s period of FRB 20201124A reported by \citet{2025arXiv250312013D} using a new method that can be applied in folding analyzes. Our conclusions are consistent with theirs. Our method can serve as a coarse approach for period searches in FRB data.


\bmhead{Acknowledgements}

This work was supported by the National Natural Science Foundation of China (grant no. 12041306).
The computation was completed on the HPC Platform of Huazhong University of Science
and Technology.

\begin{table}[htbp]
\centering
\caption{Candidate periods $P$ for different MJD observations. Notice the reported significant periods 1.706015 s on MJD
59310 and 1.707972 s on MJD 59347 by Du et al. \cite{2025arXiv250312013D} are also found in this work, with minor difference.}
\begin{tabular}{ll}
\hline
MJD & Candidate periods $P$ (s) \\
\hline
59308 & 2.167515, 4.577481, 6.377455 \\
59309 & 6.706646 \\
59310 & 1.682501, 1.707501 \\
59311 & 1.567525, 1.992523, 2.062523, 3.162520 \\
59312 & 1.812513, 4.262485 \\
59313 & 0.727512, 1.117511, 1.182510, 2.282506, 2.402506, 5.717494, 7.152488 \\
59314 & 1.147714, 3.327642, 9.547438 \\
59315 & 4.332499 \\
59316 & 6.072507, 8.932496 \\
59318 & 8.007457 \\
59319 & 1.437507, 1.812505, 2.202502, 3.042495, 3.502492, 3.997488, 7.497461 \\
59320 & 0.852504, 1.087504, 2.567501, 3.177500, 6.477494 \\
59321 & 4.167459, 8.477401, 9.217391 \\
59322 & 2.257551, 4.522531, 9.702486 \\
59323 & 3.072638, 4.347606 \\
59324 & 2.812503, 9.872493 \\
59325 & 1.782498 \\
59326 & 1.462529, 2.737507, 2.882504, 3.282497, 3.897487, 6.402443, 6.967433, 7.677421, 8.227412, 8.852401 \\
59327 & 2.547493, 3.602481, 7.952433 \\
59328 & 1.967501, 4.972498 \\
59329 & 1.997559, 9.012373 \\
59330 & 1.617543, 1.992541, 6.582517, 9.287502 \\
59331 & 2.787505 \\
59334 & 3.337522, 7.167461, 7.277460 \\
59336 & 3.857503, 6.142493 \\
59337 & 1.417784, 2.682734, 5.337628 \\
59338 & 8.992499 \\
59339 & 1.477580 \\
59340 & 1.322518, 4.962467 \\
59341 & 1.052496, 2.667446, 5.822349, 6.727321, 8.667261 \\
59343 & 1.172731, 3.897626, 4.382608, 7.797478, 9.267421 \\
59344 & 3.582509, 3.737509, 4.727507, 5.552505, 6.612503, 8.887498 \\
59345 & 2.712493, 8.192444 \\
59346 & 3.177558, 3.817551, 4.277546 \\
59347 & 1.707542, 8.411829 \\
59348 & 2.462551 \\
59349 & 1.252537, 1.777532, 3.477514, 3.682512 \\
59350 & 1.172493, 4.437472, 7.347453, 7.912449 \\
59351 & 5.047505, 8.547499, 9.942497 \\
59352 & 2.482535, 3.017531, 3.287529, 4.562520, 7.247500, 7.742497, 8.467492, 8.752489, 9.437484 \\
59353 & 2.287508, 3.152507, 6.932503, 7.472502, 9.002501 \\
59354 & 1.057509, 6.267475, 6.632473 \\
59357 & 6.747548 \\
59360 & 5.137360 \\
\hline
\end{tabular}
\label{tab:candidate_periods}
\end{table}


\bibliography{sn-bibliography}

\end{document}